\documentclass[aps,prl,twocolumn,superscriptaddress,longbibliography,amsmath,amssymb]{revtex4-2}
\usepackage{xcolor}
\usepackage{graphicx}
\usepackage{units}
\usepackage[frenchb]{babel}
\usepackage[utf8]{inputenc}
\usepackage{amssymb,amsmath}
\begin{document}
\title{Surface recombination and out of plane diffusivity of free excitons in hexagonal boron nitride}

\author{S\'ebastien Roux}
\affiliation{Laboratoire d'Etude des Microstructures, ONERA-CNRS, Universit\'e Paris-Saclay, BP 72, 92322 Ch\^atillon Cedex, France}
\affiliation{Universit\'e Paris-Saclay, UVSQ, CNRS, GEMaC, 78000, Versailles, France}
\author{Christophe Arnold}
\affiliation{Universit\'e Paris-Saclay, UVSQ, CNRS, GEMaC, 78000, Versailles, France}
\author{Etienne Carr\'e}
\affiliation{Laboratoire d'Etude des Microstructures, ONERA-CNRS, Universit\'e Paris-Saclay, BP 72, 92322 Ch\^atillon Cedex, France}
\affiliation{Universit\'e Paris-Saclay, UVSQ, CNRS, GEMaC, 78000, Versailles, France}
\author{Eli Janzen}
\affiliation{ Tim Taylor Department of Chemical Engineering, Kansas State University Manhattan, KS 66506, USA}
\author{James H. Edgar}
\affiliation{ Tim Taylor Department of Chemical Engineering, Kansas State University Manhattan, KS 66506, USA}
\author{Camille Maestre}
\affiliation{Laboratoire des Multimat\'eriaux et Interfaces, UMR CNRS 5615, Univ Lyon \\ Universit\'e Claude Bernard Lyon 1, F-69622 Villeurbanne, France}
\author{B\'erang\`ere Toury}
\affiliation{Laboratoire des Multimat\'eriaux et Interfaces, UMR CNRS 5615, Univ Lyon \\ Universit\'e Claude Bernard Lyon 1, F-69622 Villeurbanne, France}
\author{Catherine Journet}
\affiliation{Laboratoire des Multimat\'eriaux et Interfaces, UMR CNRS 5615, Univ Lyon \\ Universit\'e Claude Bernard Lyon 1, F-69622 Villeurbanne, France}
\author{Vincent Garnier}
\affiliation{Laboratoire MATEIS, UMR CNRS 5510, Univ Lyon, INSA Lyon,  F-69621 Villeurbanne, France}
\author{Philippe Steyer}
\affiliation{Laboratoire MATEIS, UMR CNRS 5510, Univ Lyon, INSA Lyon,  F-69621 Villeurbanne, France}
\author{Takashi Taniguchi}
\affiliation{Research Center for Materials Nanoarchitectonics, National Institute for Materials Science,  1-1 Namiki, Tsukuba 305-0044, Japan}
\author{Kenji Watanabe}
\affiliation{Research Center for Electronic and Optical Materials, National Institute for Materials Science, 1-1 Namiki, Tsukuba 305-0044, Japan}
\author{Annick Loiseau}
\email{annick.loiseau@onera.fr}
\affiliation{Laboratoire d'Etude des Microstructures, ONERA-CNRS, Universit\'e Paris-Saclay, BP 72, 92322 Ch\^atillon Cedex, France}
\author{Julien Barjon}
\email{julien.barjon@uvsq.fr}
\affiliation{Universit\'e Paris-Saclay, UVSQ, CNRS, GEMaC, 78000, Versailles, France}

\date{\today}
\begin{abstract}
We present a novel experimental protocol using cathodoluminescence measurements as a function of the electron incident energy to study both exciton diffusion in a directional way and surface exciton recombination. Our approach overcomes the challenges of anisotropic diffusion and the limited applicability of existing methods to the bulk counterparts of 2D materials. The protocol is then applied at room and at cryogenic temperatures to four bulk hexagonal boron nitride crystals grown by different synthesis routes. The exciton diffusivity depends on the sample quality but not on the temperature, indicating it is limited by defect scattering even in the best quality crystals. The lower limit for the diffusivity by phonon scattering is 0.2 cm$^{2}$.s$^{-1}$. Diffusion lengths were as much as 570 nm. Finally, the surface recombination velocity exceeds 10$^{5}$ cm$^{2}$.s$^{-1}$, at a level similar to silicon or diamond. This result reveals that surface recombination could strongly limit light-emitting devices based on 2D materials.
\end{abstract}

\maketitle 
 
\subsection{I - Introduction}
In most 2D semiconductors, a surprising increase in the luminescence efficiency appears when reducing the thickness at the atomic level, especially between bilayer and monolayer crystals where the bandgap undergoes its indirect to direct transition \cite{Mak2010, Tonndorf2013, Zhao2013}. However, in the specific case of hexagonal boron nitride (hBN), the bulk crystal exhibits a high radiative efficiency up to 50\% in cathodoluminescence (CL) \cite{Schue2019}, with a short radiative lifetime of ~ 30 ns \cite{Roux2021}, which arises from the high compacity of its indirect excitons. Conversely, the luminescence intensity for hBN thin films \cite{Schue2016} is dramatically decreased : there is no CL signal for less than 6 layers. The luminescence of the monolayer hBN free exciton has only recently been observed, thanks to extremely long photoluminescence experiments \cite{Elias2019, Rousseau2021}. The underlying reasons for the decrease in intensity with the thickness of the hBN layers are yet to be fully understood and elucidated. It will be explored in this work in terms of surface effect and exciton diffusion.

Often ignored in 2D materials, surfaces play a well-known role in 3D semiconductors in limiting light emission efficiency in standard optoelectronic devices. The rate of surface recombination of free charge carriers is governed by the physical properties of the defect states at the surface, so that it could be reduced by using passivation strategies \cite{Nolte1990, Hsu1992, Fitzgerald1968}. It is quantitatively characterised by a parameter called the surface recombination velocity, $s$, which typically ranges between 1 and 10$^{6}$ cm.s$^{-1}$ depending on the material and its surface termination. \cite{Nolte1990, Jastrzebski1975, Fitzgerald1968, Kozak2012}. In hBN, due to the sp$^{2}$ orbital hybridisation, all chemical bonds are satisfied between atoms on the same atomic plane, so there are no dangling bonds at the surface. Considering this, the surface recombination rate for free carriers should be low in hBN compared to sp$^{3}$ hybridised semiconductors with dandling bounds \cite{Nolte1990}. However, strong surface effects are suggested by the low luminescence efficiency of hBN crystals of a few atomic layers \cite{Schue2016}. In hBN, and more generally in 2D materials, surface recombinations are still unexplored and remain an open question.

Surface recombination and the associated luminescence losses are promoted by the diffusion of charge carriers injected into the volume of a crystal towards its surfaces. In the context of lamellar materials, such as the 2Ds, the out-of-plane diffusivity is the key parameter for understanding the surface recombination properties. Furthermore, understanding exciton diffusion in 2D semiconductors is of strategic importance for the development of the new "excitronic" devices based on the control of exciton flux, which has attracted attention with the advent of 2D heterostructures \cite{High2008, Butov2017, Jauregui2017, Unuchek2018}. However, experiments and devices have only focused on the in-plane exciton diffusion in 2D monolayers of transition metal dichalcogenides \cite{Cadiz2018, Kumar2014, Uddin2020}.

The study of exciton diffusion can be carried out by PL experiments using various pump-probe techniques \cite{Heller1996, Lawson1982}, or by CL experiments, which benefit from a better control of the excitation volume, down to tens of $nm^3$, by analysing the transfer of excitons from a localised excitation to a physical probe \cite{Zarem1989, Ino2008, Hocker2016, Kenkre1981, Barjon2012, Ino2008, Nogues2014}. To properly analyse such experiments, one must decouple the combined effects of (i) in-plane diffusion, (ii) out-of-plane diffusion, (iii) surface recombinations, and (iv) inhomogeneous in-depth energy deposition. Due to the anisotropy and the still unexplored surface recombination properties of 2D materials, the use of traditional protocols for their out-of-plane exciton diffusion poses significant challenges. 

This work presents a novel experimental protocol in CL that provides a powerful tool for investigating the out-of-plane diffusion and the surface recombination properties of a bulk crystal using a 1D excitation geometry. The method is applicable to all types of semiconductors, including those with anisotropic diffusion, and does not require prior knowledge of the surface recombination properties. After introducing the principle of the experiment, a theoretical analysis of diffusion is given, providing an analytical description of experimental results. The method is then applied to four bulk hBN samples at room and cryogenic temperature, allowing the evaluation of out-of-plane diffusion length, diffusion constant, surface recombination velocity, and their dependence on the crystal quality and temperature.

\subsection{II - Analytical model for a 1D diffusion experiment in CL}

\begin{figure}[h!]
\includegraphics[scale=0.4]{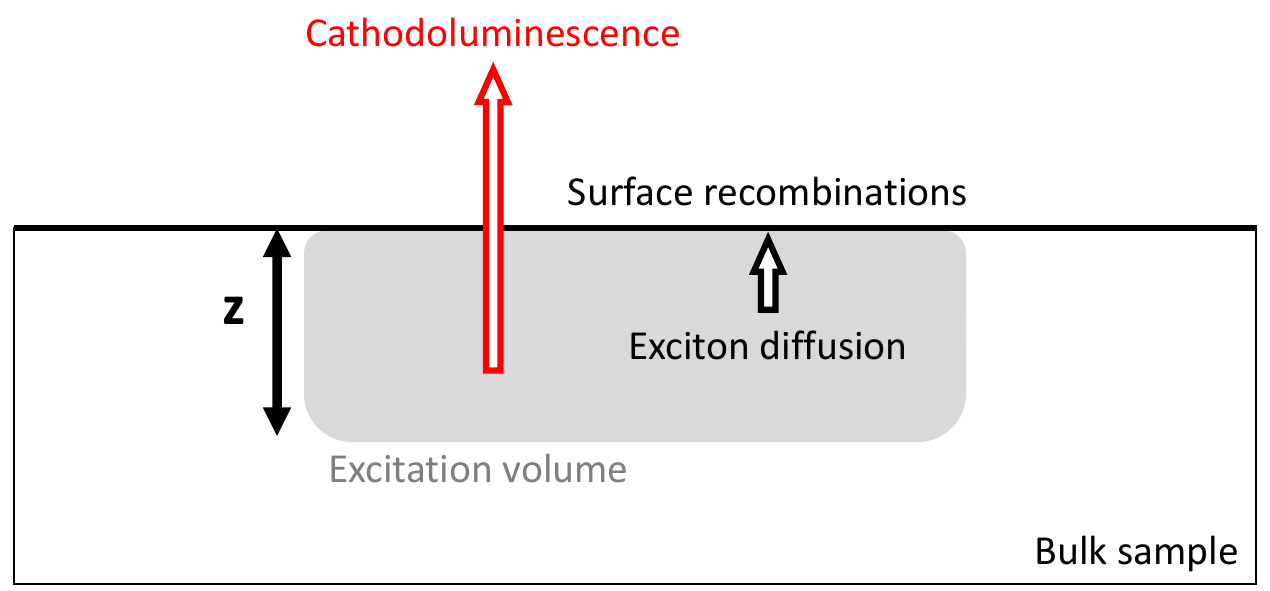}
\caption{Scheme of the CL experiment in a 1D geometry to analyse the out-of-plane exciton diffusion in the bulk crystal and the recombination properties of the surface.}
\label{F1}
\end{figure}

We first present the principle of the CL experiment. As illustrated in Figure \ref{F1}, a 1D diffusion geometry is obtained by defocusing significantly the incident electron beam. The CL intensity is measured as a function of the excitation depth, which is modulated by varying the acceleration voltage of the incident electrons. At low excitation depths, surface recombination effects lead to a reduction of the CL intensity and the key point is to use the recombinant surface as a probe for exciton diffusion. 
The quantitative analysis of such an experiment requires the consideration of two essential aspects: (i) the development of a simple and quantitative description of the excitation depth as a function of the acceleration voltage, and (ii) the establishment of an analytical model that explains the decrease in CL intensity at low excitation depths, taking into account the surface effect and in relation to the relevant physical parameters. We first address theoretically these two issues.

The free exciton \footnote{Assuming strongly bounded excitons, i.e. negligible free carrier concentrations.} concentration $n(z)$ that is established under steady-state generation $\mathcal{G}$ in a semiconductor crystal results from the diffusion equation studied here in a 1D symmetry :
\begin{equation} D\frac{\partial^2 n}{\partial z^2} - \frac{n}{\tau} + \mathcal{G} =0
\end{equation} 

$D$ is the exciton diffusivity along the $z$ axis, and its determination is one of the goals of the CL experiments. $\tau$ is the exciton lifetime in the bulk of the crystal, i.e. far from the surface. It is measured by time-resolved CL (TRCL) experiments at high voltage as detailed in a previous work \cite{Roux2021}.

We consider here a semi-infinite crystal surrounded by vacuum for $z<0$. The surface recombinations at $z=0$ are described as a boundary condition of the diffusion equation, with the surface recombination velocity $s$ :

\begin{equation}
D\left [\frac{\partial n}{\partial z}\right]_{z=0}= s \: n(0)
\end{equation} 

Dealing with the generation $\mathcal{G}$ due to an excitation spread in depth is a rather difficult mathematical problem. The solution for a surface source located in a plane at the depth $z=z'$ was formulated in 1955 by van Roosbroeck \cite{Roosbroeck1955}, where the depth distribution of particles in the crystal is defined separately on each side of the generation plane:

\begin{equation}
f_{z', a}(z) = \left\{
\begin{array}{ll}
\frac{1}{2L}e^{-\frac{z'}{L}}\left(e^{\frac{z}{L}}+ae^{-\frac{z}{L}}\right)\text{ if } z \leqslant z' \\
\frac{1}{2L}e^{-\frac{z}{L}}\left(e^{\frac{z'}{L}}+ae^{-\frac{z'}{L}}\right)\text{ if } z \geqslant z'
\end{array}
\right. 
\label{fza}\end{equation} 

where $L=\sqrt{D\tau}$ is the diffusion length along the $z$ axis. An interpretation of this result with virtual image sources is as follows: the recombinant surface at $z=0$ might be viewed as a semi-reflecting mirror for the real sources in the crystal with a reflection coefficient $a=\frac{1-S}{1+S}$. In this expression $S=\frac{s}{v_D}$ is the reduced recombination velocity, i.e. normalised by the diffusion velocity $v_D = \frac{L}{\tau}=\sqrt{\frac{D}{\tau}}$. The limit cases are: $S$=+$\infty$ ($a$=-1) for a surface with infinite recombination velocity and $S$=0 ($a$=1) assimilated to a total reflection.

Since the diffusion equation is linear, an arbitrary in-depth profile of exciton generation $g(z)$ normalised such as $\int_0^{+\infty} g(z)dz = 1$ results in an exciton concentration distribution that is written here \footnote{$g(z)=\delta(z-z')$ for a plane source in $z=z'$}:

\begin{equation}
n(z) = \frac{G \tau}{\mathcal{A}}\int_0^{+\infty} f_{z', a}(z) g(z')dz' \label{nz}
\end{equation}

This expression indicates how excitons are distributed in the semiconductor crystal by the combined effects of diffusion from a given excitation profile and recombinations at the crystal surface. The pre-factor contains the experimental parameters of a CL experiment. $\mathcal{A}$ is the area of the electron beam impinging on a 3D crystal. Experimentally one should ensure that $\mathcal{A} \gg L^2$ to assume a 1D diffusion. For instance, the electron beam spot should be sufficiently defocused to be much larger than the in-plane diffusion length (see Figure \ref{F1}). $G$ (s$^{-1}$) is the electron-hole generation rate in the semiconductor. $G=\frac{(1-f) V i}{\langle E_{eh} \rangle }$ is a simple function of the acceleration voltage, $V$, the beam current, $i$, the retrodiffusion factor, $f$, and the average electron-hole energy, $\langle E_{eh} \rangle$ required for the formation of an electron-hole pair. $\langle E_{eh} \rangle \approx 3 E_g$ is a commonly used relation as a function of the bandgap energy $E_g$ when $\langle E_{eh} \rangle$ is not known experimentally for the material of interest \cite{Klein1968}.

The recombinations at the crystal surface at $z=0$ are responsible for exciton losses and \textit{in fine} for a decrease of their luminescence intensity $I$. Instead of integrating Eq. \ref{nz} over the crystal volume to evaluate the total number of steady-state excitons in the crystal, the quantity of lost excitons $N'$ is calculated more directly \cite{Yacobi2013} considering the diffusion flux at $z=0$ :
\begin{equation}
N'= D\left [\frac{\partial n}{\partial z}\right]_{z=0} \mathcal{A}\tau 
\end{equation}

which gives the following result after calculation:
\begin{equation}
\frac{N'}{N} = \frac{(1-a)}{2} \int_0^{+\infty} e^{-\frac{z'}{L}} g(z')dz' 
\end{equation}

where $N=G\tau$ is the steady-state exciton population in a crystal without surface losses ($s=0$). The luminescence intensity $I$ is proportional to the total number of excitons present in the crystal under steady-state excitation, via the constant radiative rate of excitons. Noting $I_0$ the luminescence intensity when surface recombination is negligible, measured experimentally in the high voltages limit, we obtain $\frac{I}{I_0} = 1 - \frac{N'}{N}$. Finally, for a constant generation rate $G$, the CL intensity evolves as:

\begin{equation}
\frac{I}{I_0} = 1- \frac{(1-a)}{2} \int_0^{+\infty} e^{-\frac{z'}{L}} g(z')dz'\label{Igenerale}
\end{equation}

\begin{figure}[h!]
\includegraphics[scale=0.22]{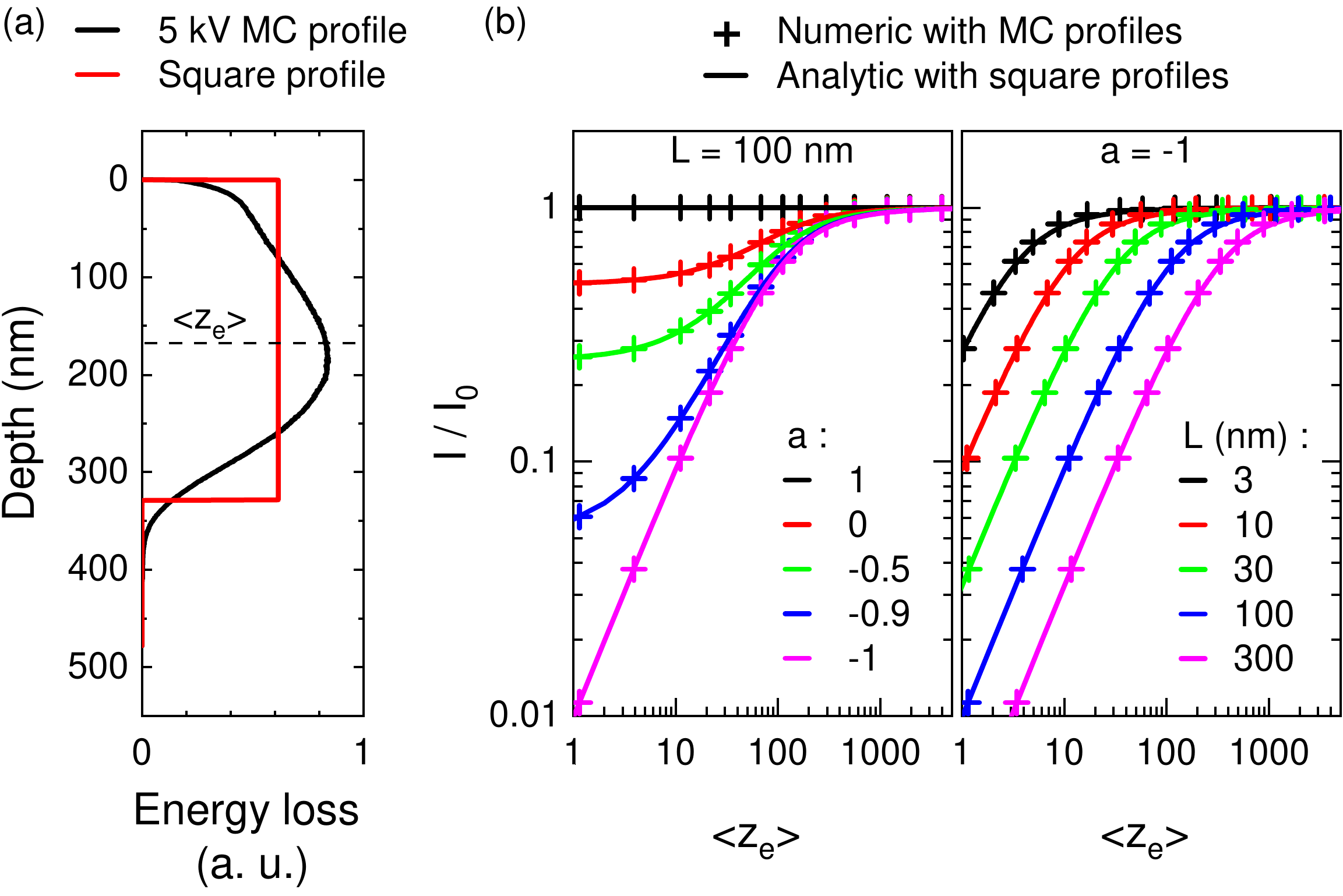}
\caption{ (a) Generation profile for a 5 kV incident electron beam in hBN calculated by MC simulations (in black), and its square approximation (in red). Both profiles share the same mean excitation depth $\langle z_{e} \rangle$ calculated by MC simulation. (b) $I / I_{0}$ as a function of $\langle z_{e} \rangle$ for a steady excitation inside the crystal. The results are obtained either numerically with discrete generation profiles from MC simulations (cross) or analytically with Eq. \ref{Ianalytique} considering the square profile.
}
\label{F2}
\end{figure}

Various analytical functions have been proposed to describe $g(z)$, the in-depth profile of excitation by an electron beam as function of its acceleration voltage \cite{Kanaya1972, Sieber2012}. Their success was limited for quantitative analysis of cathodoluminescence experiments because their validity is restricted to limited ranges of acceleration voltage and material density. Today, Monte Carlo (MC) algorithms are preferred since they provide simulations of electron energy losses in a simple and relatively accurate way \cite{Drouin2007}. As shown in the Supplementary Materials, the in-depth distribution of the generation rate can be assessed numerically with Monte-Carlo simulations as well as the acceleration voltage dependence of the mean excitation depth $\langle z_{e} \rangle = \int_0^{+\infty} z g(z)dz$. An example of calculated profile for a 5kV acceleration voltage is shown in Figure \ref{F2} (a) together with the corresponding value of $\langle z_{e} \rangle$.

Considering its shape, this profile can be approximated by a square profile also shown in Figure \ref{F2}(a) and defined by a depth distribution $g(z)=\sqcap(z)$, with a constant generation rate down to an electron stopping depth $R =2 \langle z_{e} \rangle$, i.e. :

$$
\sqcap(z) = \left\{
\begin{array}{ll}
1/R & \mbox{if } 0\leqslant z \leqslant R \\
0 & \mbox{if } z > R
\end{array}
\right.
$$

Using this simpler depth distribution is of key interest since its provides an analytical expression of $\frac{I}{I_0}$ involving three parameters namely $a$, $L$ and $\langle z_{e} \rangle$ :

\begin{equation}
\frac{I}{I_{0}}=1-\frac{(1-a)}{2} \;\frac{1-e^{-\frac{2 \langle z_{e} \rangle}{L}}}{\frac{2 \langle z_{e} \rangle}{L}}
\label{Ianalytique}
\end{equation}

Figure \ref{F2}(b) depicts the variations of $\frac{I}{I_0}$ as a function of $\langle z_{e} \rangle$ for different values of $a$ and $L$ as generated from the analytical equation Eq. \ref{Ianalytique}. They are compared with numerical calculations of $\frac{I}{I_0}$ obtained by discretising the integral term in Eq. \ref{Igenerale} to inject the generation profiles $g(z)$ obtained from MC simulations.

This comparison attests that an excellent agreement is found between analytical and numerical variations. This remarkable result shows that the precise shape of the generation profile $g(z)$ has a negligible effect on the luminescence intensity, while the mean excitation depth $\langle z_{e} \rangle$ turns out to be the critical parameter for the interpretation of these CL experiments. This observation holds true even for GaN, which has a significantly higher density ($6.15 g.cm^{-3}$) than hBN ($2.18 g.cm^{-3}$), resulting in drastically different excitation profiles (for further details see the Supplementary Materials). Thus, it turns out that the analytical description of Eq. \ref{Ianalytique} can be applied to a wide range of semiconductor materials, as soon as the CL reabsorption by the crystal remains negligible, which is the case in indirect band gap semiconductors for instance.
\footnote{The present model is inspired by the early works of Wittry et al. in 1967. \cite{Wittry1967} Despite the title of their paper, ``Measurement of Diffusion Lengths in Direct Gap Semiconductors by Electron Beam Excitation'', the present work rather applies to indirect semiconductors, since the efficient luminescence reabsorption in direct bandgap semiconductors requires careful consideration when evaluating luminescence intensity.}

Figure \ref{F2}(b) also reveals that the parameters $a$ and $L$ have distinct influences on the $\frac{I}{I_{0}}(\langle z_{e} \rangle)$ curve. In particular, the parameter $a$ determines the shape of the curve at low excitation depths, while the parameter $L$ governs the position of the curve along the horizontal axis. This observation provides the guidance for the analysis of experimental measurements. Provided to vary the acceleration voltage over a sufficiently large range, the analytical description of Eq. \ref{Ianalytique} has the ability to accurately fit experimental measurements of the CL intensity independently with reliable values of the diffusion length $L$ and the parameter $a$ which quantifies the level of surface recombinations. Further, combining these experiments with measurements of the bulk exciton lifetime $\tau$ using TRCL at high voltage as introduced above gives access to the evaluation of the diffusivity $D$ and the surface recombination velocity $s$ which are the key physical quantities of interest.

\subsection{III - Experimental results for free excitons in hBN : analysis and discussion}

We now turn to the experiments. Four bulk hBN crystals were investigated. One was grown at the Kansas State University using an Atmospheric Pressure High Temperature process \cite{Liu2018, Li2020, Kubota2007} from Ni/Cr solvent and natural boron isotop content (APHT sample). Two crystals were synthesised at the NIMS laboratory using a High Pressure High Temperature route \cite{Watanabe2004, Taniguchi2007} (samples HPHT1 and HPHT2). And the last crystal was grown in the LMI laboratory using a Polymer Derived Ceramic method \cite{Li2018, Li2020b} (PDC sample). The HPHT samples are widely recognised in the 2D material scientific community as reference hBN materials. Prior to analysis, the samples were cleaved to obtain a clean, uncontaminated surface, and then mounted on conducting doped silicon wafers. A conductive bond was established on one edge of the crystal using Ag paint. This mounting allows an efficient charge evacuation during electron beam irradiation. The thickness of the crystals was measured to be greater than 10 $\mu$m using a mechanical profilometer.

Cathodoluminescence spectra were collected out using a Scanning Electron Microscope (SEM) JEOL7001F equipped with a Horiba Jobin-Yvon CL system optimised for UV spectroscopy, as described in detail in reference \cite{Schue2019}. To avoid non-linear effects such as exciton-exciton annihilation \cite{Plaud2019}, a low excitation power of about 1 $\mu$W was employed. To achieve a 1D geometry configuration (shown in Figure \ref{F1}), the electron beam was significantly defocused, resulting in an excitation spot size of 3.5 $\mu$m. The excitation depth was controlled by varying the electron acceleration voltage from 500 V to 20 kV. For each acceleration voltage, the CL intensity $I$ measurements were normalised by the excitation power $P=iV \cdot (1-f)$, taking into account slight variations ($\pm$10\%) in the incident electron beam power ($iV$) at each acceleration voltage $V$, as well as the voltage dependence of the electron backscattering factor $f$ (see Supplementary Materials). The uncertainty in the CL intensity was estimated to be 20\%. Finally, time-resolved CL was performed using a custom-built beam blanker installed in the SEM column, achieving a temporal resolution of 100 ps (details of the setup can be found in \cite{Roux2021}).
\begin{figure}[h!]
\includegraphics[scale=0.2]{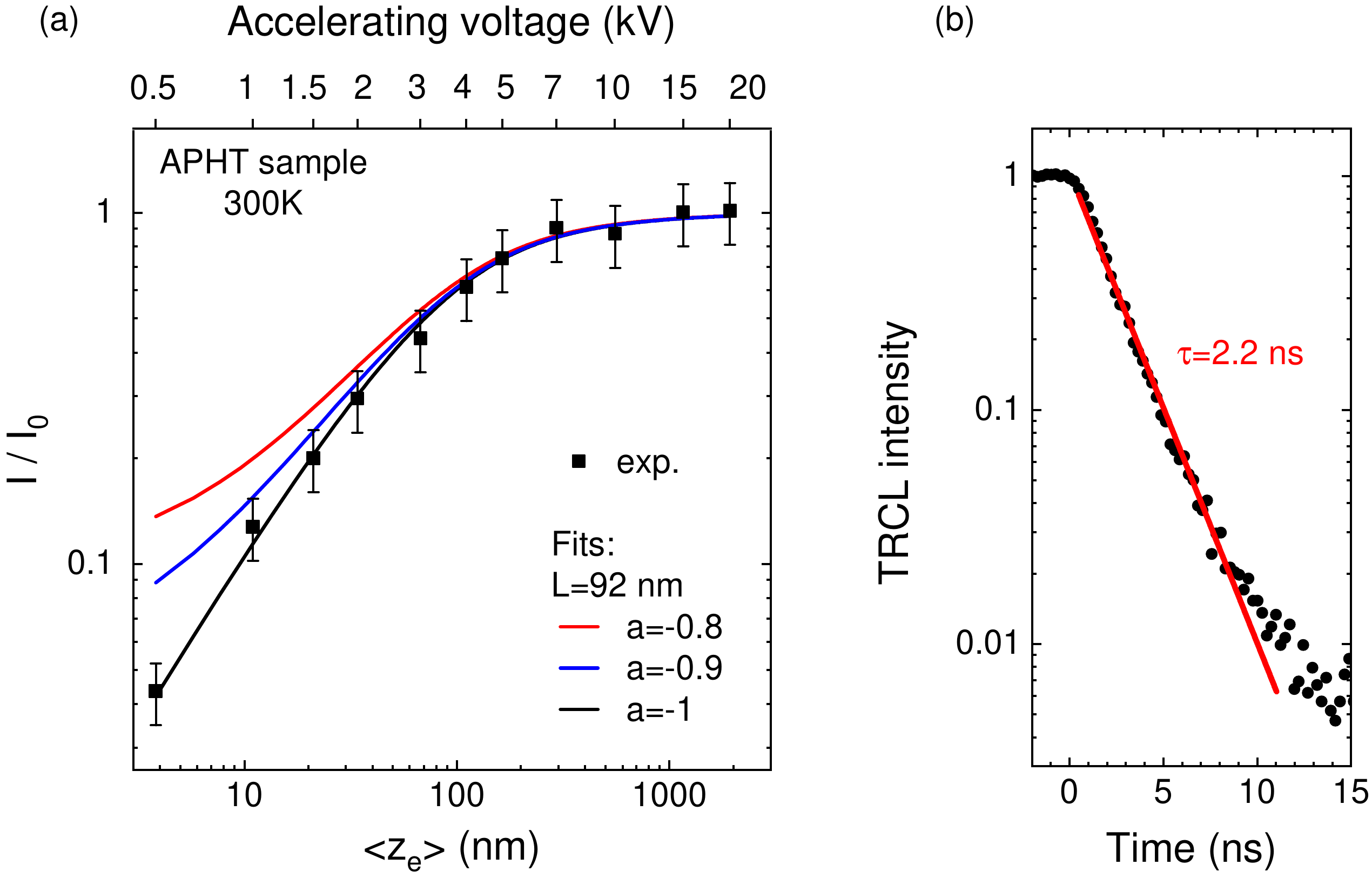}
\caption{(a) CL intensity $I/I_0$ as a function of the acceleration voltage and the corresponding mean excitation depth $\langle z_e \rangle$ (see Supplementary Materials). The analytical equation (\ref{Ianalytique}) is depicted for a diffusion length $L$ of 92 nm and various values of the parameter $a$. (b) Temporal decay of the free exciton luminescence measured in the same area after the interruption of a 15 kV electron beam. The bulk lifetime of the free exciton $\tau$ is determined by fitting the data with an asymptote.}
\label{F3}
\end{figure}
Figure \ref{F3} presents the first application of the protocol to the APHT sample. In Figure \ref{F3}(a), the CL intensity $I/I_0$ is plotted as a function of the mean excitation depth $\langle z_e \rangle$ as extracted from Monte Carlo simulations. At high excitation depths, the intensity plateaus indicate that excitons are generated too far from the top surface to cause significant surface effects. The decrease in luminescence intensity at low excitation depths is attributed to surface recombinations enhanced by exciton diffusion towards the top surface. Fit of the experimental data by the equation \ref{Ianalytique} is discussed from Figure \ref{F3}(a). Taking into account the uncertainties, the 95\% confidence interval for $L$ is between 84 nm and 100 nm, and the best fit is with $L = 92$ nm as shown in the Figure \ref{F3}(a). This Figure also illustrates that only a few points at low excitation depths contribute to the determination of the parameter $a$. Equation \ref{Ianalytique} is displayed for $L = 92$ nm and $a = -1$, $a = -0.9$, and $a = -0.8$. Reasonable agreement is found for $a < -0.9$, while higher values lead to a poorer fit, establishing an upper limit of $-0.9$ for $a$.

These results make it possible to evaluate and discuss the diffusivity $D$ and the surface recombination velocity $s$ by starting with the latter. The surface recombination velocity $s = \frac{L}{\tau} \frac{1 - a}{1 + a}$ is known from $L$, $\tau$, and $a$. The exciton lifetime $\tau$ is measured independently by time-resolved CL with an uncertainty of 10\%, following the procedure described in \cite{Roux2021} and shown in Figure \ref{F3}(b). Knowing the upper limit of $a$, a lower limit for the surface recombination velocity $s$ is deduced, yielding $s > 7 \times 10^4$ cm/s for the APHT sample.

The high value of $s$ is very robust as it could be confirmed from the analysis of data obtained from the different samples and at various temperatures (as shown in Tables \ref{t1} and \ref{t2}), leading to refinement of its lower limit to $10^5$ cm/s at 6 and 300 K. This is definetely a surprinsingly high value, approaching the highest levels reported around $10^6$ cm/s for 3D semiconductors such as in GaAs \cite{Jastrzebski1975} and Diamond \cite{Kozak2012}. A possible explanation for this result could be the presence of defects at the hBN surface. While it is challenging to completely rule out this possibility, note that the surfaces studied in this work were clean and freshly cleaved, minimising potential contamination. In addition, there is no change in the shape of the spectra with the depth of excitation, suggesting a defect concentration homogeneous in depth and a surface free of contamination, as shown in the spectra in the appendices. Another possible explanation relates to the surface exciton states that have been identified theoretically \cite{Paleari2018} and experimentally \cite{Schue2019} in hBN, and which are approximately 100 meV below the bulk hBN exciton. These surface states suggest that the hBN surface acts as a sink for excitons, with a sufficient energy depth to capture them efficiently at room temperature, consistently with the high surface recombination rate observed in hBN. 

Whatever its origin, the importance of surface recombination in hBN allows us to understand the difficulty of measuring the luminescence of its atomic layers. In the presence of such a high surface recombination rate, the luminescence of hBN crystals made of a few atomic layers is inevitably very low, as experimentally measured by L. Schué et al. \cite{Schue2016}. Contrary to initial assumptions, the surfaces of 2D materials can also limit the efficiency of their luminescence, emphasising the need to passivate them for use in optoelectronic devices. Recent studies on atomic layers of transition metal dichalcogenides (TMDs) have demonstrated that the luminescence quantum yields only reach unity when the material is properly passivated \cite{Lien2019, Amani2016}. The increase in quantum efficiency following passivation indicates that non-radiative recombinations occur at the TMD-air surface. Similar passivation techniques should be considered to mitigate non-radiative recombinations at the hBN surface, therby enabling enhanced luminescence intensity and improving the potential of hBN for optoelectronic applications.

\begin{figure}[h!]
\includegraphics[scale=0.22]{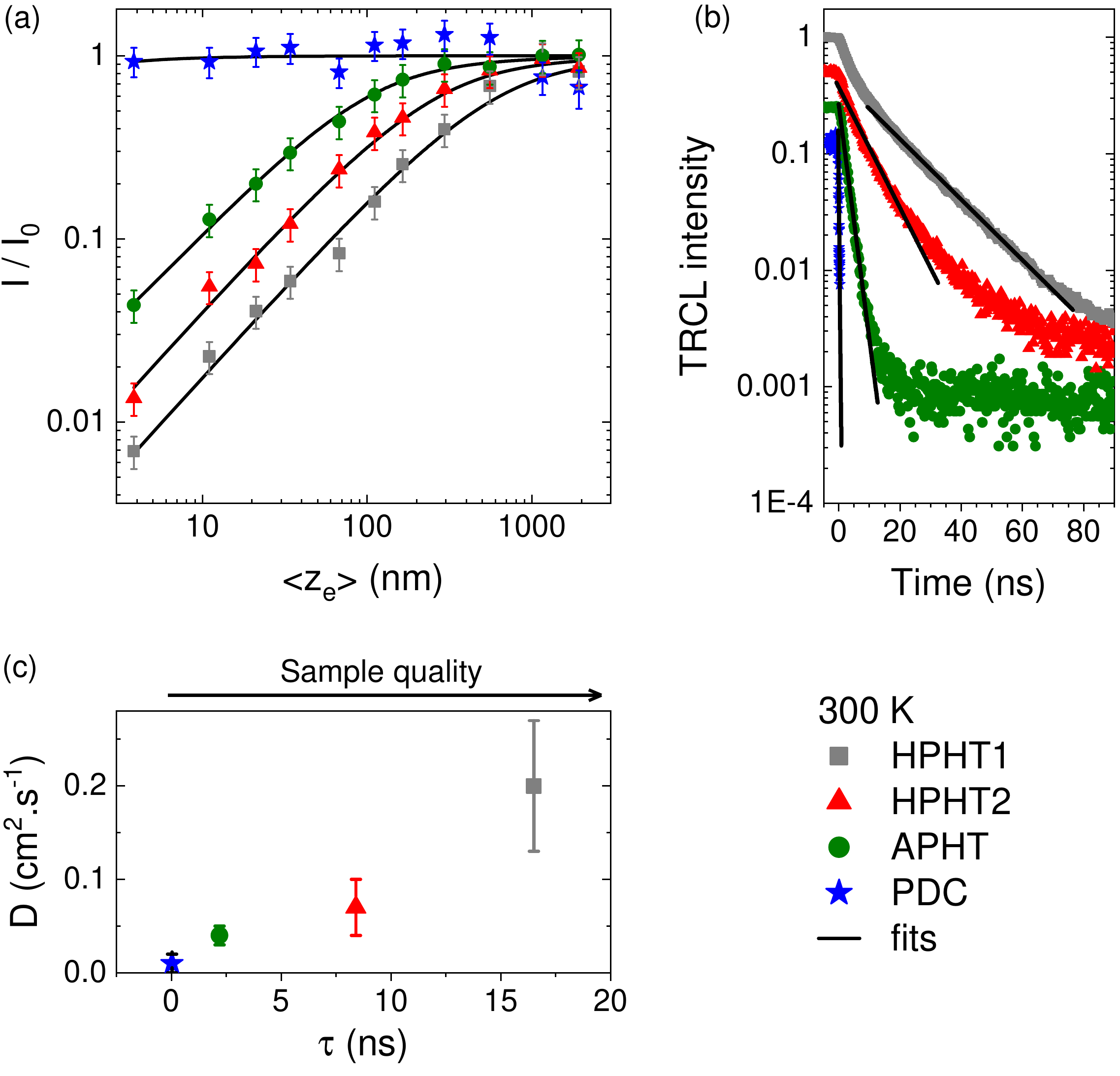}
\caption{(a) $I / I_{0}$ as function of the mean excitation depth $\langle z_{e} \rangle$, and (b) temporal decays of the free exciton luminescence after interruption of a 15 kV steady excitation at $t$=0, recorded on HPHT1, HPHT2, APHT and PDC samples. The out-of-plane diffusion length $L$ is deduced by fitting the equation (\ref{Ianalytique}) in (a), while the bulk exciton lifetime $\tau$ is measured by TRCL in (b). (c) Diffusivity as function of the exciton lifetime $\tau$ for the 4 samples. Note that $\tau$ is used to quantitatively compare the global quality of the samples.
}
\label{F4}
\end{figure}

We now focus on the diffusion properties. The measured diffusion lengths $L$ and the exciton lifetimes $\tau$ on the four samples at 300 K are depicted in Figure \ref{F4} (a-b). In the case of the PDC sample, the exciton lifetime is too short to be measured directly by TRCL. However, using the linearity relationship between $\tau$ and the CL efficiency given in reference \cite{Roux2021}, the exciton lifetime $\tau$ on this sample was estimated as 0.03 $\pm$ 0.015 ns from the measurement of a 0.1\% CL efficiency (details of the measurement in reference \cite{Schue2019}). For all other samples, $\tau$ is measured directly by TRCL with a 10\% uncertainty. Considering the uncertainty of the CL intensity measurement, the 95\% confidence intervals of $L$ are always found within $\pm$ 15\%. Finally, the diffusivity $D$ is extracted considering the relation $D=L^{2}/\tau$ and taking into account the uncertainty on $L$ and $\tau$. The results are reported in the Table \ref{t1}. 
As explained in reference \cite{Roux2021}, the higher the quality of the sample, the longer the exciton lifetime is. In the Figure \ref{F4} (c), the exciton lifetime $\tau$ is used to quantitavely compare the globale quality of the 4 samples. The figure exhibits a variation of the diffusivity $D$ of more than one order of magnitude between the different samples and clearly demonstrates that the diffusivity is related to the defects present in the samples. 

\begin{figure}[h!]
\includegraphics[scale=0.21]{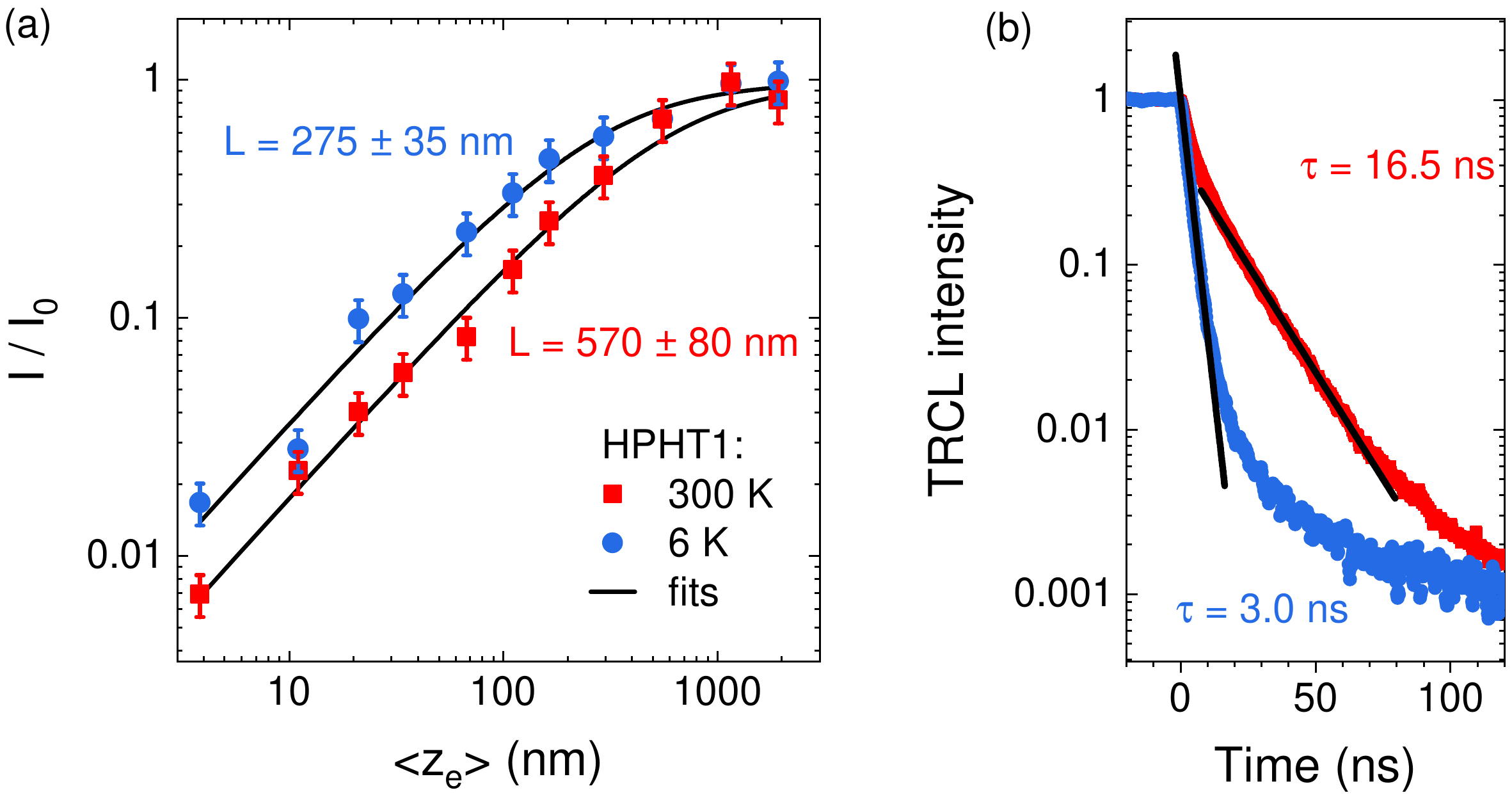}
\caption{(a) $I / I_{0}$ as function of the mean excitation depth $\langle z_{e} \rangle$ during continuous excitation and (b) temporal decays of the free exciton luminescence after interruption of the 15 kV electron beam, recorded on the HPHT1 sample at 6 and 300 K.
}
\label{F5}
\end{figure}

\begin{table}[h!]
\caption{Results of the CL experiments on the four bulk hBN samples at 300 K : the out-of-plane exciton diffusion length $L$, the parameter $a$ and the exciton lifetime $\tau$. An estimate of the diffusivity $D=L^{2}/\tau$ and a lower bound of the surface recombination velocity $s=\frac{L}{\tau} \frac{1-a}{1+a}$ are derived.}
\begin{ruledtabular}
\begin{tabular}{c|ccccc}
  & $L$ (nm) & $a$ & $\tau$ (ns) & $s$ (cm/s) & $D$ (cm$^{2}$/s) \\
\hline
HPHT1 & 570 & $<$ -0.98 & 16.5 & $>$3*10$^{5}$ & 0.20 $\pm$ 0.07 \\
\hline
HPHT2 & 246 & $<$ -0.96 & 8.4 & $>$1*10$^{5}$ & 0.07 $\pm$ 0.03 \\
\hline
APHT & 92 & $<$ -0.9 & 2.2 & $>$7*10$^{4}$ & 0.04 $\pm$ 0.01 \\
\hline
PDC & $<$ 5 & - & 0.03 & - & $<$ 0.02 \\
\end{tabular}
\end{ruledtabular}
\label{t1}
\end{table}
Figure \ref{F5} presents the measurements of the diffusion length $L$ and exciton lifetime $\tau$ on the best sample (HPHT1) at both 6 K and 300 K using our experimental protocol. The corresponding results are summarised in Table \ref{t2}. The diffusivity is 0.25 $\pm$ 0.08 cm$^{2}$/s at 6 K and 0.20 $\pm$ 0.07 cm$^{2}$/s at the room temperature. The difference is not significant given the uncertainties of the measurement. In the case of intrinsic diffusion limited by phonon scattering, the diffusion constant at cryogenic temperatures is expected to be several orders of magnitude higher than at room temperature \cite{Morimoto2015}. Our results indicate that even in the highest-quality HPHT1 sample, exciton diffusion is not limited by phonon scattering but rather by defects scattering. This is consistent with impurity concentrations in oxygen and carbon in the order of 10$^{18}$ cm$^{-3}$ or below measured in the reference HPHT samples \cite{Taniguchi2007, Onodera2019} a value which remains high compared to ultrapure semiconductors (10$^{11}$-10$^{13}$ cm$^{-3}$ typically) where the intrinsic diffusivity is reached \cite{Morimoto2015}. 

\begin{table}[h!]
\caption{Results of the CL experiments at 6 K and 300 K on the HPHT1 sample.}
\begin{ruledtabular}
\begin{tabular}{c|ccccc}
T° (K) & $L$ (nm) & $a$ & $\tau$ (ns) & $s$ (cm/s) & $D$ (cm$^{2}$/s) \\
\hline
300 K & 570 & $<$ -0.98 & 16.5 & $>$3*10$^{5}$ & 0.20 $\pm$ 0.07 \\
\hline
6 K & 275 & $<$ -0.95 & 3 & $>$3*10$^{5}$ & 0.24 $\pm$ 0.08 \\
\end{tabular}
\end{ruledtabular}
\label{t2}
\end{table}

An estimate of the lower limit of the intrinsic diffusivity in hBN is provided by the highest value measured on our sample : $D$ $>$ 0.2 cm$^{2}$/s. Few values of the exciton diffusion constant are available in the literature for comparison, since excitons in standard semiconductors are weakly bounded and generally dissociated at room temperature. Among them, the diffusion constant is 5 cm$^{2}$/s in diamond at 300 K \cite{Morimoto2015}. In organic semiconductors where the diffusion is limited by exciton hoping from site to site, $D$ is found in the range  10$^{-6}$ - 0.01 cm$^{2}$.s$^{-1}$  \cite{Mikhnenko2015}. The lower bound found for $D$ in hBN indicates that the hypothesis of a hoping diffusion mechanism can be safely discarded in this material, which has a diffusion constant typical of a standard semiconductor.

\subsection{IV - Conclusion}
In conclusion, we have developed a new CL protocol thanks to which the out-of-plane exciton diffusion length $L$, the diffusivity $D$ and the surface recombination velocity $s$ have been recorded for the first time in a 2D semiconductor. It consists in recording the emitted light as function of the acceleration voltage over a large range and using a widely defocused beam to be in 1D geometry. By a detailed analysis of the 1D diffusion equation applying to these CL experiments, we show that the emitted light in the presence of surface recombinations and diffusion is not affected by the in-depth profile of the exciton generation function, but by the mean excitation depth $\langle z_{e} \rangle$ identified as the key parameter for our experiment. A simple analytical function is derived to describe the experimental results and to estimate the physical quantities in turn. The application of the protocol to four bulk hBN samples reveals diffusion lengths $L$ up to 570 nm at 300 K in the best crystal. From temperature dependent experiments, the exciton diffusivity is shown to be limited by defects scattering. Its lower limit for the intrinsic diffusivity is 0.2 cm$^{2}$/s, similar to the value expected for a standard semiconductor. Finally, the hBN surface recombination velocity is extremely high, at the level of silicon or diamond. This surprising result explains the low intensity of hBN thin films, and reveals that exciton recombination at surfaces could be a strong limitation for light-emitting devices based on 2D materials.

\begin{acknowledgments}
The research leading to these results has received funding European Union's Horizon 2020 research and innovation program under grant agreements No 785219 (Graphene Core 2) and No 881603 (Graphene Core 3). Support for the APHT hBN crystal growth comes from the Office of Naval Research, Award No. N00014-20-1-2474. K.W. and T.T. acknowledge support from the JSPS KAKENHI (Grant Numbers  21H05233 and 23H02052) and World Premier International Research Center Initiative (WPI), MEXT, Japan.
\end{acknowledgments}

%

\end{document}


\title{Annexes : Surface recombinations and out of plane diffusivity of free excitons in hexagonal Boron Nitride}
\maketitle

\subsection*{Annexe I : Monte-Carlo simulations to evaluate the in-depth energy losses of incident electrons}

\begin{figure}[h]
\includegraphics[scale=0.21]{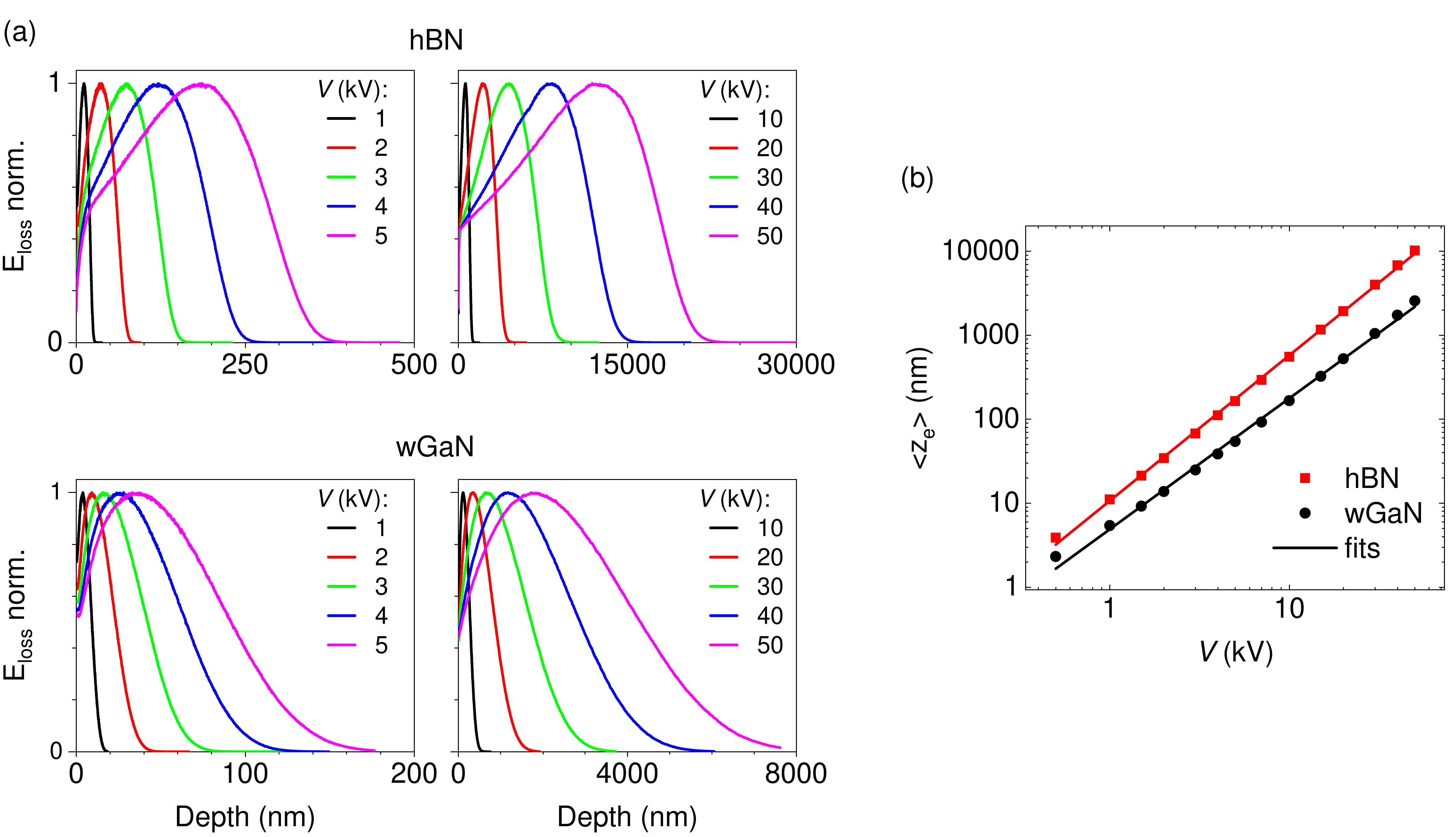}
\caption{(a) Electron energy losses as function of the depth, in hBN and wGaN, for different acceleration voltage $V$ of the incident electron beam. (b) Mean excitation depth $\langle z_{e} \rangle$ as function of the acceleration voltage $V$. The results are fitted by power laws.}
\label{FA1}
\end{figure}

Monte Carlo (MC) simulations using Casino V2.42 software \cite{Drouin2007} have been used to evaluate the in-depth distribution of electron energy losses in two different materials: hexagonal boron nitride (hBN, density 2.18 g.cm$^{-3}$) and wurtzite gallium nitride (wGaN, density 6.15 g.cm$^{-3}$). The simulations were performed to support the experimental study on hBN and to test the analytical model on a denser material (wGaN). The simulations followed the path of 10$^6$ incident electrons for each acceleration voltage ($V$) in the respective materials. The energy losses were calculated as a function of depth, and the mean excitation depth $\langle z_{e} \rangle = \int_0^{+\infty} z g(z)dz$ was determined (Figure \ref{FA1} and Table \ref{tA1}). The results reveal that the in-depth profiles of energy losses present asymmetry and significant differences between hBN and wGaN, as shown in the figures.

The excitation power of the sample can be determined using the relation $P = iV \cdot (1 - f)$, where $i$ and $V$ are the current and the acceleration voltage of the incident electron beam, and $f$ corresponds to the loss due to backscattering of incident electrons. The parameter $f$ can be calculated from the Monte Carlo simulations (results in table \ref{tA1}). The excitation power $P$ is used to normalise the cathodoluminescence (CL) intensity (see main text).

\begin{table}[h!]
\caption{Mean excitation depth $\langle z_{e} \rangle$ and backscattering factor $f$, calculated from the Monte-Carlo simulations on hBN and wGaN for different acceleration voltages $V$.}
\begin{ruledtabular}
\begin{tabular}{cc|cccccccccccccc}
\multicolumn{2}{c|}{$V$ (kV)}& 0.5 & 1&1.5&2&3&4&5&7&10&15&20&30&40&50 \\
\hline
\multirow{2}{*}{hBN}&$<z_{e}> (nm)$&3.9&11&21&34&68&111&164&295&555&1160&1930&3980&6790&10200 \\
\cline{2-16}
 &f&0.1&0.08&0.07&0.06&0.06&0.05&0.05&0.05&0.04&0.04&0.04&0.04&0.03&0.03\\
\hline
\multirow{2}{*}{wGaN}&$<z_{e}> (nm)$& 2.3&5.4&9.2&14&25&38&54&92&167&324&525&1050&1730&2570 \\
\cline{2-16}
&f&0.27&0.27&0.26&0.26&0.25&0.25&0.25&0.25&0.24&0.24&0.24&0.24&0.23&0.22

\end{tabular}
\end{ruledtabular}
\label{tA1}
\end{table}

Figure 1 (b) presents the mean excitation depth $\langle z_{e} \rangle$ as a function of the incident energy $E_{0}$ within the range of 500 V to 50 kV. The data follow power laws, namely $\langle z_{e} \rangle_{hBN} = 10.67 \; V^{1.73}$ and $\langle z_{e} \rangle_{wGaN} = 4.89 \; V^{1.56}$, where $\langle z_{e} \rangle$ is given in nm and $V$ is in kV. It is worth noting that these power laws align well with the empirical relations used in the past, such as the Kanaya-Okayama equation \cite{Kanaya1972}, which describes the stopping depth of electrons with a power dependence of 5/3.

\subsection*{Annexe II : Validation of the analytical model on hBN and GaN}

\begin{figure}[h]
\includegraphics[scale=0.26]{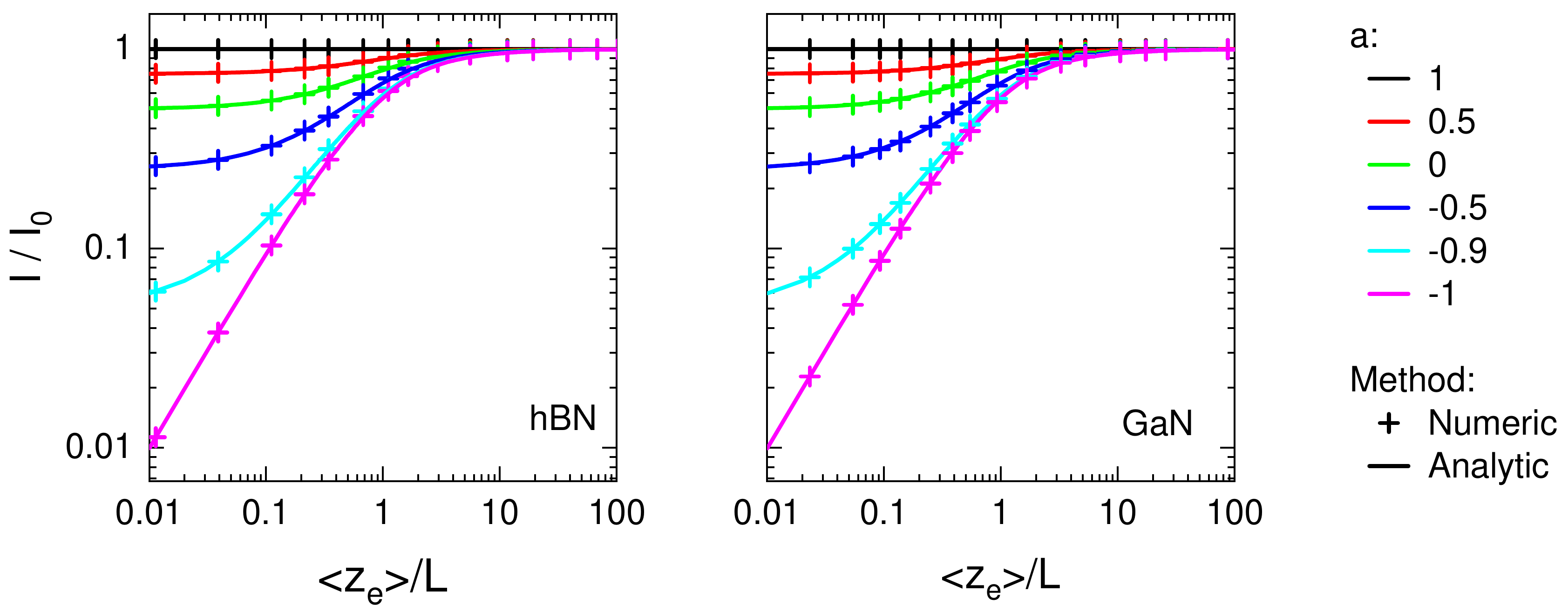}
\caption{$I / I_{0}$ as a function of $\langle z_{e} \rangle$/L. The results are obtained numerically using the generation profiles from the MC simulations (cross) on hBN and GaN, and analytically with the square generation profile approximation (see Figure 2 (a) in the main text).}
\label{FA2}
\end{figure}

Figure \ref{FA2} depicts a theoretical calculation of $I / I_{0}$ as a function of $\langle z_{e} \rangle / L$, where $\langle z_{e} \rangle$ represents the mean excitation depth and $L$ denotes the diffusion length. The calculation takes into account the realistic in-depth excitation profiles obtained from Monte Carlo (MC) simulations (refer to Figure \ref{FA1}), as well as simplified square profiles (Figure 2 (a) in the main text). The comparison is made between two materials with different densities: hBN (2.18 g/cm$^{-3}$) and wGaN (6.15 g/cm$^{-3}$). The abscissa is dimensionless thanks to a normalisation by L, in order to explore all the couples [$L$; $a$].

The figure demonstrates a remarkable agreement between the results obtained using equation (8), which solely considers the mean excitation depth $\langle z_{e} \rangle$, and the results obtained from Monte Carlo (MC) simulations incorporating the in-depth excitation profile through discretised equation (7). This agreement holds true for both light materials such as hBN and heavier materials like wGaN, and regardless of the values of the parameters $a$ and $L$. This validation demonstrates that $\langle z_{e} \rangle$ is the key parameter to interpret our experiment and that the analytical expression is applicable to fit the experimental results for a large panel of semiconductors and to deduce $L$ and $a$.

\subsection*{Annexe III : Spectra recorded on the HPHT2 sample}
\begin{figure}[h]
\includegraphics[scale=0.35]{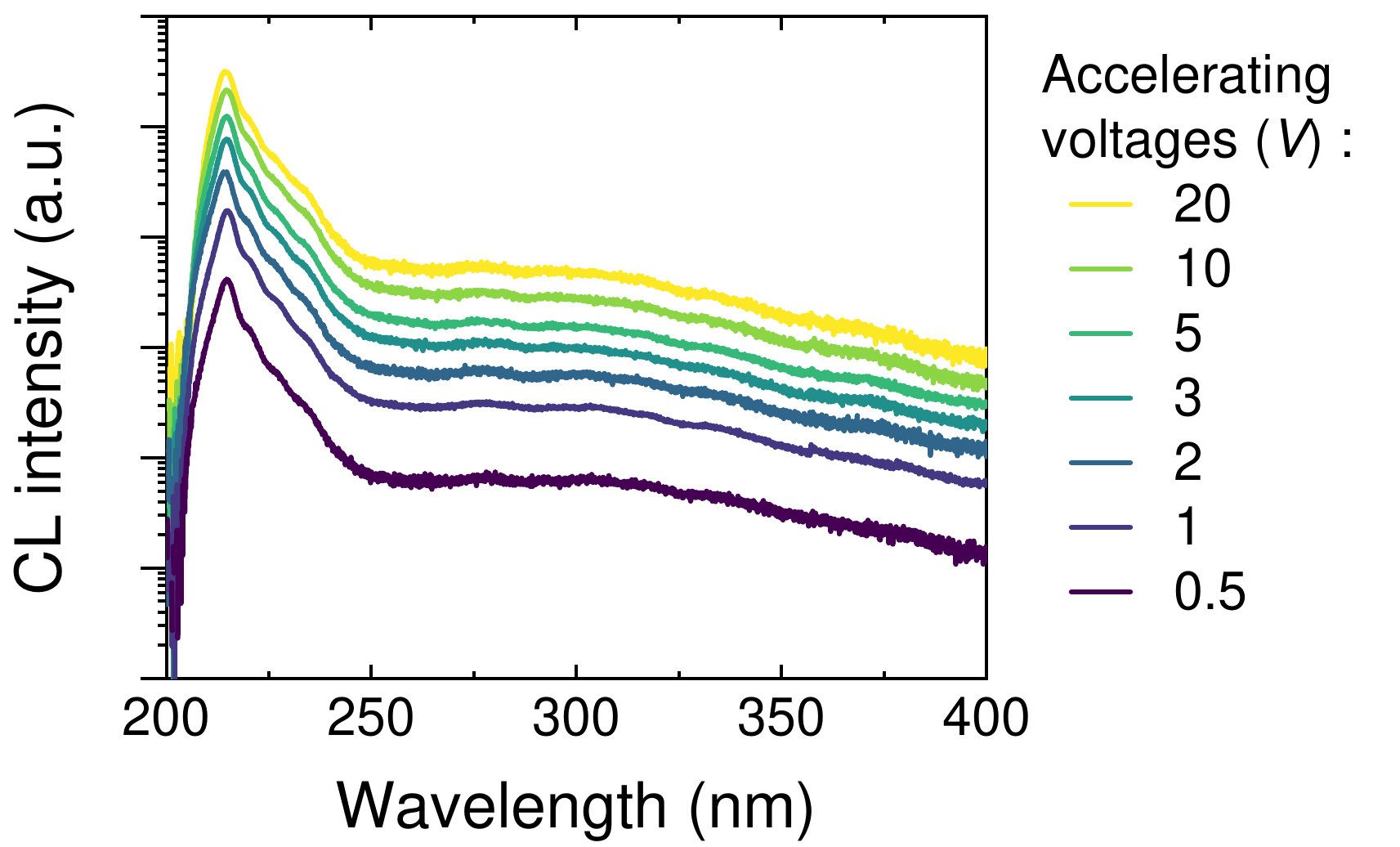}
\caption{Spectra acquired on HPHT2 sample at 300 K, for different acceleration voltages and with a constant excitation power of 1 µW.}
\label{FA3}
\end{figure}

The figure \ref{FA3} shows the spectra obained on the HPHT2 sample at 300 K and for different acceleration voltages ranging between 500 V and 20 kV. The spectra are corrected by the spectral response of the detector, following the procedure described in reference \cite{Schue2019}. They are dominated by the free exciton luminescence which appears at 300 K as a broad band between 200 and 240 nm with a maximum at 215 nm \cite{Schue2019}. Defect-related luminescence appears at higher wavelengths \cite{Schue2016Characterization}, here with a low intensity. The intensity $I/I_{0}$ used in the main text is extracted from the free exciton luminescence intensity measured on these spectra.

As observed in the figure \ref{FA3}, the shape of the spectra is identical regardless of the acceleration voltage, which means that it does not depend on the depth of excitation. This suggests that the crystal is extremely homogeneous in depth and presents a surface free of contamination. The decrease in intensity observed at a low acceleration voltage is then attributed to the recombination of excitons at the surface of the hBN-air rather than to the presence of impurities on the surface.


%